# Theoretical Investigation of Subluminal Particles Endowed with Imaginary Mass

Luca Nanni

University of Ferrara, 44122 Ferrara, Italy
luca.nanni@edu.unife.it

**Abstract** In this article, the general solution of the tachyonic Klein-Gordon equation is obtained as a Fourier integral performed on a suitable path in the complex $\omega$-plane. In particular, it is proved that under given boundary conditions this solution does not contain any superluminal components. On the basis of this result, we infer that all possible spacelike wave equations describe the dynamics of subluminal particles endowed with imaginary mass. This result is validated for the Chodos equation, used to describe the hypothetical superluminal behaviour of neutrino. In this specific framework it is proved that the wave packet propagates in spacetime with subluminal group velocities and that for enough small energies it behaves as a localized wave.

## 1. Introduction

The physics of tachyons is an intriguing and fascinating subject that has attracted the attention of physicists in both the pre-relativistic [1-3] and post-relativistic epochs [4-10]. However, the majority of the scientific community still considers it a speculative theory, and few are the efforts aimed at finding the experimental evidence that could candidate it to be an extension of the current Standard Model [11]. A comprehensive overview of the most relevant experiments dealing with the superluminal behaviour of waves and particles is detailed in the references [9,12-15]. If considered in the framework of quantum mechanics, tachyons exhibit completely unexpected behaviours [16-18]. In particular, half-integer free fermions endowed with imaginary mass behave like wave trains propagating with subluminal group velocities. The aim of this study is to verify whether this result can be generalized for any particle regardless of its spin. The basic idea is that the individual components of spinor wave functions, which are solutions of the tachyonic wave equations, must satisfy the tachyonic Klein-Gordon (TKG) equation as well [19]. Finding a general solution of the TKG equation without spacelike components means proving that any solution of the wave equations of particles with negative mass squared represents a wave packet that propagates with subluminal group velocity. As proposed by Salesi, these particles are called pseudo-tachyons (PTs), namely particles travelling with slower-than-light velocity but fulfilling the tachyonic energy-momentum relation $E^2 = p^2c^2 + \mu^2c^4$, where $\mu = im$ denotes the imaginary mass. The obtained result will be validated for the Chodos equation [20] which describes the behaviour of a ½-spin tachyon using a more different analytical approach than that discussed in the reference [18].

## 2. Preliminary Notions on the Fourier Transforms in the Complex Plane

To facilitate the reading of § 3 of this article, here is a brief review on the theory of the Fourier transform in a complex plane [21].
Let be $f(x)$ an integrable function; its Fourier transform is defined as:

$$\mathcal{F}[f(x)] = \tilde{f}(\omega) = \int_{-\infty}^{\infty} f(x)e^{i\omega x}dx, \tag{1}$$

while the inverse Fourier transform is:

$$\mathcal{F}^{-1}[\tilde{f}(\omega)] = \frac{1}{2\pi} \int_{-\infty}^{\infty} \tilde{f}(\omega)e^{-i\omega x}d\omega. \tag{2}$$

Definitions (1) and (2) can be generalized by allowing $\omega$ to take complex values. To this end let suppose that $f(x)$ grows at most exponentially at infinity, i.e. $f(x) = \mathcal{O}(e^{c|x|})$ as $x \to \pm\infty$ and $c > 0$. Moreover, we split up $f(x)$ as follows:

$$f(x) = f_+(x) + f_-(x) : f_+(x) = 0 \; \forall x < 0 \; and \; f_-(x) = 0 \; \forall x > 0 \tag{3}$$

The Fourier transform of the component $f_+(x)$ is:

$$\tilde{f}_+(\omega) = \int_0^{\infty} f_+(x)e^{iRe(\omega)x}e^{-Im(\omega)x}dx. \tag{4}$$

It is evident that:

$$|\tilde{f}_+(\omega)| \leq \int_0^{\infty} |f_+(x)|e^{-Im(\omega)x}dx, \tag{5}$$

and the integral converges provided $Im(\omega) > c$. Therefore, $\tilde{f}_+(\omega)$ exists and is holomorphic for $Im(\omega) > c$.

Next, we need extend the Fourier inversion theorem to recover $f_+(x)$ from $\tilde{f}_+(\omega)$. To this end, let $F_+(x) = f_+(x)e^{-\alpha x}$, where $\alpha > c$ so that $[\tilde{F}_+(\omega)] = \tilde{f}_+(\omega + i\alpha)$ exists and is holomorphic for $Im(\omega) > c - \alpha$, in particular for $\omega \in \mathbb{R}$, since $\alpha > c$. Thus, we can apply the Fourier transform inversion theorem obtaining:

$$F_+(x) = \frac{1}{2\pi} \int_{-\infty}^{\infty} \tilde{F}_+(\omega)e^{-i\omega x}d\omega \; \Rightarrow \; f_+(x)e^{-\alpha x} = \frac{1}{2\pi} \int_{-\infty}^{\infty} \tilde{f}_+(\omega + i\alpha)e^{-i\omega x}d\omega$$

$$\Rightarrow f_+(x) = \frac{1}{2\pi} \int_{-\infty}^{\infty} \tilde{f}_+(\omega + i\alpha)e^{-i(\omega+i\alpha)x}d\omega. \tag{6}$$

The final integral corresponds to integration along a horizontal contour in the complex $\omega$-plane:

$$f_+(x) = \frac{1}{2\pi} \int_{-\infty+i\alpha}^{\infty+i\alpha} \tilde{f}_+(\omega) e^{-i\omega x} d\omega. \tag{7}$$

Suppose $\tilde{f}_+(\omega)$ can be continued below $Im(\omega) = c$, so that it is holomorphic in some region $\Omega_+ \supset \{\omega : Im(\omega) > c\}$ except for singularities at $\omega = a_1, a_2, \cdots$. By the deformation theory, the inversion contour $\Gamma_+ = \{x + i\alpha : -\infty < x < \infty\}$ may be deformed into $\Omega_+$ provided it passes above the singularities of $\tilde{f}_+(\omega)$. Since the singularities of $\tilde{f}_+(\omega)$ are below the inversion contour, for $x < 0$ we can close the contour in at $+i\infty$. This gives the expected result that $f_+(x) = 0$ for $x < 0$. For $x > 0$, we would need to close the contour in $Im(\omega) < 0$, picking up the contributions from the singularities in $\tilde{f}_+(\omega)$ and giving a nonzero value of $f_+(x)$.

The same procedure works for $f_-(x)$ with everything upside down, namely $\tilde{f}_-(\omega)$ can be continued above $Im(\omega) = -c$, so that it is holomorphic in some region $\Omega_- \supset \{\omega : Im(\omega) < -c\}$ except for singularities at $\omega = b_1, b_2, \cdots$. If there is a non-empty overlap region $\Omega = \Omega_+ \cap \Omega_- \setminus \{a_i\} \cup \{b_j\}$, where $\{a_i\}$ are the singularities of $\tilde{f}_+(\omega)$ and $\{b_j\}$ are the singularities of $\tilde{f}_-(\omega)$, then the Fourier transform of $f(x)$ is defined as:

$$\tilde{f}(\omega) = \tilde{f}_+(\omega) + \tilde{f}_-(\omega), \tag{8}$$

for $\omega \in \Omega$. Moreover, if $\Gamma_+$ and $\Gamma_-$ can be deformed into the same contour $\Gamma \subset \Omega$, with $\Gamma$ above the singularities of $\tilde{f}_+(\omega)$ and below the singularities of $\tilde{f}_-(\omega)$, then:

$$f(x) = \frac{1}{2\pi} \int_\Gamma \tilde{f}(\omega) e^{-i\omega x} d\omega. \tag{9}$$

What is discussed in this section represents the method that will be implemented below to solve the TKG equation.

## 3. Dynamics of Wave Packets in the TKG equation

The TKG equation can be obtained by quantising the tachyonic energy-momentum relation using the operators $E \to i\hbar\, \partial/\partial t$ and $p \to -i\hbar c \nabla$:

$$\left( \hbar^2 \frac{\partial^2}{\partial t^2} - \hbar^2 c^2 \nabla^2 - m^2 c^4 \right) \psi(\mathbf{r}, t) = 0. \tag{10}$$

For free scalar particles Eq. (10) can be solved by Fourier transform in time, which allows decomposing the wave function $\psi(\mathbf{r}, t)$ in its monochromatic temporal components, using the initial conditions given for $\psi(\mathbf{r}, t)|_{r=0}$ and $\partial \psi(\mathbf{r}, t)/\partial \mathbf{r}|_{r=0}$. To simplify the following discussion, we suppose that the motion of the particle takes place along the $z$-axis. The solution is a wave packet represented by the following inverse Fourier transform:

$$\psi(z,t) = \frac{1}{2\pi} \int_\Gamma \left[ A(\omega)e^{iK(\omega)z} + B(\omega)e^{-iK(\omega)z} \right] e^{-i\omega t} d\omega, \tag{11}$$

where $\Gamma$ is a path in the $\omega$-plane. The quantity $\omega$ is defined as $\omega = \pm E/\hbar$, and for a tachyon is always a real quantity except when $E = \pm\mu c^2$ (this last statement will be clarified in § 4 where we calculate the group velocity of the tachyonic wave packet). In fact:

$$E = \pm\gamma\mu c^2 = \pm\left(1 - \frac{u^2}{c^2}\right)^{-1/2} imc^2 = \pm\left(\frac{u^2}{c^2} - 1\right)^{-1/2} mc^2, \tag{12}$$

where $\gamma$ is the tachyonic Lorentz factor and $u > c$ is the classical tachyonic velocity. Returning to Eq. (11), the term $K(\omega)$ is the dispersion relation whose characteristic equation is:

$$K^2(\omega) = \frac{\omega^2 - \omega_0^2}{c^2} \quad where \quad \omega_0^2 = -m^2c^4/\hbar^2. \tag{13}$$

Eq. (13) is obtained from Eq. (10) upon factorisation $\psi(z,t) = \varphi(z)e^{-i\omega t}$. The aim is to prove that Eq. (1) admits a general solution in which no spacelike components arise. In other words, we want to prove that this solution is completely analogous to the one that solves the ordinary KG equation, where $\omega$ is always real, and tends to the solution of the d'Alembert wave equation as the mass energy tends to zero. We point out that the spacelike components are those associated with complex values of $\omega$, and therefore of $K(\omega)$.

The coefficients $A(\omega)$ and $B(\omega)$ are calculated once the following boundary conditions have been set [21]:

$$f(t) = \psi(0,t) \quad and \quad f'(t) = \partial\psi(z,t)/\partial z|_{z=0}. \tag{14}$$

As we will see, these conditions determine the choice of the path $\Gamma$ of integration. Using boundary conditions (14), we get [21]:

$$\begin{cases} A(\omega) = \frac{1}{2}\int_{-\infty}^{\infty}\left[f(t) - i\frac{f'(t)}{K(\omega)}\right]e^{i\omega t}dt \\ B(\omega) = \frac{1}{2}\int_{-\infty}^{\infty}\left[f(t) + i\frac{f'(t)}{K(\omega)}\right]e^{i\omega t}dt \end{cases} \tag{15}$$

The computation of integrals (15) can be simplified if we consider $f(t)$ as a semi-infinite wave trains with sharp edge. In other words, we can set a time $t_0$ such that $\forall t < t_0 \Rightarrow f(t) = 0$. This is equivalent to state that at the time $t_0$ the wave is created. However, in order to ensure the convergence of the integral (11), as $t \to t_{0^+}$ the function $f(t)$ tends to zero. From now on we assume $t_0 = 0$. Under this assumption, being $f(t)$ and $f'(t)$ infinite wave trains with a sharp edge, it is necessary that $\omega$ runs on a path equivalent to the line $(-\infty + i\alpha, \infty + i\alpha)$ where $\alpha > 0$, in order to obtain the convergence of the integrals (15).

Let us now set the following further condition:

$$B(\omega) = 0 \Rightarrow f'(t) = \frac{1}{2\pi} i \int_\Gamma K(\omega) \mathcal{F}(f) e^{-i\omega t} d\omega, \tag{16}$$

where $\mathcal{F}(f)$ is the Fourier transform of the function $f(t)$. Using the condition (16) the first of integrals (15) becomes:

$$A(\omega) = \int_{-\infty}^{\infty} f(t) e^{i\omega t} dt = \mathcal{F}(f). \tag{17}$$

Substituting integral (17) in Eq. (11) we get:

$$\psi(z,t) = \frac{1}{4\pi} \int_\Gamma \mathcal{F}(f) e^{iK(\omega)z - i\omega t} d\omega. \tag{18}$$

Since $f(t)$ is a semi-infinite sharp-edged wave train, it can be written explicitly as:

$$f(t) = \Theta(t) e^{-i\omega|_{t_0} t}, \tag{19}$$

where $\Theta(t)$ is the Heaviside function. The Fourier transform of function (19) is [22]:

$$\mathcal{F}(f) = i \frac{1}{\omega - \omega|_{t_0}}. \tag{20}$$

Substituting function (20) in the Eq. (18) we obtain the explicit form of the kernel:

$$\psi(z,t) = \frac{1}{2\pi} \int_\Gamma \frac{i}{\omega - \omega|_{t_0}} e^{iK(\omega)z - i\omega t} d\omega. \tag{21}$$

To solve integral (21) we need finding the integration path $\Gamma$. In order to satisfy the initial conditions at $z = 0$, all singularities of the Fourier coefficient $A(\omega)$ must lie under the path $\Gamma$ (as discussed in § 2), so that closing $\Gamma$ in the upper half-plane the integral (11) vanishes as $t < t_0$. This is the mathematical representation of the fact that since the wave function is a semi-infinite train with a sharp edge, the integrals (15) give rise to singularities on the real axis of the complex plane, except for the two branch points at $\omega = \pm i\omega_0$. Only non-physical expressions for $f(t)$ and $f'(t)$ (namely, not suitable for a wave function) can produce singularities in the upper half-plane. The two branch points come from the dispersion relation:

$$K(\omega) = \pm \frac{1}{c}(\omega^2 + \omega_0^2)^{1/2} = \pm \frac{1}{c}[(\omega + |\omega_0|)(\omega - |\omega_0|)]^{1/2}. \tag{22}$$

There are two possible kinds of cuts: the segment $(-i\omega_0, i\omega_0)$ or the pair of half-line $(i\omega_0, i\omega_0 + \infty)$ and $(-i\omega_0, -i\omega_0 + \infty)$. Since the path $\Gamma$ must lies above the real axis we have to study how it behaves when it meets the cut. It could go around the cut or run parallel to the real axis, both of these choices are mathematically correct. However, the second choice would lead to arise of superluminal components given by the upper cut, and the group velocity associated to the wave function will be greater than the speed of light. The criterion for discriminating the choice of the integration path is that at the limit $|\omega_0| \to 0$ the dispersion relation $K(\omega)$ tends to $\omega/c$, which represents the typical dispersion relation of a wave packet that satisfies the ordinary wave equation, which does not admit tachyonic components. This condition excludes the cuts given by the pair of half-line $(i\omega_0, i\omega_0 + \infty)$ and $(-i\omega_0, -i\omega_0 + \infty)$ and leads to the right dispersion relation $K(\omega) = |\omega^2 + \omega_0^2|^{1/2}/c$ for $\omega > 0$ and $K(\omega) = -|\omega^2 + \omega_0^2|^{1/2}/c$ for $\omega < 0$. The wave function (21) has a completely general form and the integration path $\Gamma$ is uniquely determined by the choice of the initial conditions (14), (16) and (19). Therefore, it is expected that whatever the tachyonic equation considered, the wave packet describing the dynamics of the particle always propagates with subluminal velocity, even if it is associated with a negative mass squared.

## 4. Validation of the Obtained Result for Half-integer Spin Tachyon

The results obtained in § 3 should be validated for each tachyonic wave equation. However, to the best of our knowledge the only tachyonic wave equation formulated for massive particles is that of Chodos, which describes the motion of a particle with half-integer spin [23]. This equation is generally used to investigate neutrinos exhibiting superluminal behaviours [24]. Therefore, we will work on this equation with the aim to obtain a solution in the form of a Gaussian wave packet from which calculating the group velocity. To this purpose, we formulate the equation of the envelope function that characterized the wave packet.

The Chodos equation is obtained from the Tanaka Lagrangian [25]. For a tachyonic particle moving along the $z$-axis this (Hermitian) Lagrangian reads:

$$\mathcal{L} = i\hbar\bar{\psi}\gamma^5\gamma^0\partial_t\psi - i\hbar c\bar{\psi}\gamma^5\gamma^3\partial_z\psi - mc^2\bar{\psi}\psi, \tag{23}$$

where, $m = |\mu|$, $\bar{\psi}_t = \psi_t^\dagger\gamma^0$ and $\gamma^5 = i\gamma^0\gamma^1\gamma^2\gamma^3$. Introducing the Lagrangian (23) in the Euler-Lagrange equations we obtain the Chodos equation:

$$(i\hbar\gamma^5\gamma^0\partial_t - i\hbar c\gamma^5\gamma^3\partial_z - mc^2)\psi = 0 \tag{24}$$

The spacelike property of Eq. (24) is given by the operator $\gamma^5$ and not by the imaginary mass [25]. We seek solution that have the form of a Gaussian wave packet:

$$\psi^\pm(z,t) = \mathcal{N}\begin{pmatrix}\varphi_1^\pm \\ \varphi_2^\pm\end{pmatrix}f^\pm(z,t)exp\{\pm i(kz - \omega^\pm t)\}, \tag{25}$$

where $\pm$ denotes the solutions with positive and negative frequencies, $\varphi_{1,2}^\pm$ are the spinor components, $f^\pm(z,t)$ is the Gaussian envelope function and $\mathcal{N}$ is the normalization constant. Before formulating the equation for the function $f^\pm(z,t)$ we need to calculate the spinor components for the particle and antiparticle states. To this purpose, we substitute the plan wave component of Eq. (25) in Eq. (24) obtaining a system of four algebraic equations by which to calculate the spinor components:

$$\begin{pmatrix} i\hbar c\partial_z - mc^2 & 0 & -i\hbar\partial_t & 0 \\ 0 & -i\hbar c\partial_z - mc^2 & 0 & -i\hbar\partial_t \\ i\hbar\partial_t & 0 & -i\hbar c\partial_z - mc^2 & 0 \\ 0 & i\hbar\partial_t & 0 & i\hbar c\partial_z - mc^2 \end{pmatrix} \begin{pmatrix} \varphi_1^+ \\ \varphi_2^+ \\ \varphi_1^- \\ \varphi_2^- \end{pmatrix} e^{\pm i(kz-\omega^\pm t)} = 0. \quad (26)$$

Multiplying Eq. (26) times its adjoint we get four TKG equations. This equation can be easily solved by Cramer rule obtaining:

$$\varphi_1^+ = \varphi_2^- = -\left(\frac{\omega}{kc+\omega_0}\right)^{1/2} \quad ; \quad \varphi_2^+ = \varphi_1^- = \left(\frac{\omega}{kc-\omega_0}\right)^{1/2}. \quad (27)$$

where $\omega_0 = mc^2/\hbar$. The spinor components of Eq. (27) can be rewritten in parametric form. To this end we set the classical tachyonic velocity as $u = \eta c$ with $\eta > 1$. Then, Eq. (27) becomes:

$$\varphi_1^+ = \varphi_2^- = -\left[\frac{(\eta^2-1)^{-1/2}}{\eta(\eta^2-1)^{-1/2}+1}\right]^{1/2} \quad ; \quad \varphi_2^+ = \varphi_1^- = \left[\frac{(\eta^2-1)^{-1/2}}{\eta(\eta^2-1)^{-1/2}-1}\right]^{1/2}. \quad (28)$$

We can now formulate the equation in $f^\pm(z,t)$ which allows us to calculate both the group velocity of the wave packet and the explicit form of the envelope function. Substituting Eq. (25) in Eq. (24) we obtain:

$$\begin{cases} \left(\frac{\partial}{\partial t} - c\frac{\varphi_1^+}{\varphi_1^-}\bigg|_{\eta_0}\frac{\partial}{\partial z} + \frac{2mc^2}{i\hbar}\frac{\varphi_1^+}{\varphi_1^-}\bigg|_{\eta_0}\right)f^+(z,t) = 0 \\ \left(\frac{\partial}{\partial t} + c\frac{\varphi_2^-}{\varphi_2^+}\bigg|_{\eta_0}\frac{\partial}{\partial z} + \frac{2mc^2}{i\hbar}\frac{\varphi_2^-}{\varphi_2^+}\bigg|_{\eta_0}\right)f^-(z,t) = 0 \end{cases}, \quad (29)$$

where $\eta_0$ is the center of the Gaussian envelope function. The numerical coefficient of the second term in Eq. (29) is just the group velocity of the wave packet [26]:

$$v_g = -c\frac{\varphi_1^+}{\varphi_1^-}\bigg|_{\eta_0} = c\frac{\varphi_2^-}{\varphi_2^+}\bigg|_{\eta_0} = c\left[\frac{\eta_0(\eta_0^2-1)^{-1/2}-1}{\eta_0(\eta_0^2-1)^{-1/2}+1}\right]^{1/2} < c \quad \forall \eta_o > c. \quad (30)$$

As expected, Eq. (30) proves that the group velocity is always lower than the speed of light, confirming the result obtained in § 3. From Eq. (30) we see that when the classical tachyonic velocity $u$ tends to infinity, i.e. when $\eta_0 \to \infty$, then the group velocity $v_g$ tends asymptotically to zero. In other words, in a reference frame moving with an infinitely large relative velocity, the wave packet is localized. This is another relevant result of this study and it allows to reinterpret the Hartman effect [27] in quantum tunnelling phenomena. This will be discussed in more detail in § 5. Furthermore, the limit $v_g \to 0$ when $\eta_0 \to \infty$ can be used to explain why the *rest mass energy* of the PT is given by $\mu c^2$. The imaginary value of this energy is due to the proper time $\tau$ of a tachyon which, as is known [28-29], is imaginary.

## 5. Discussion

In this study we proved that, once appropriate initial conditions have been set,

is always possible to find a general solution of the TKG equation in which no superluminal components arise and that gives the same results of the ordinary wave equation at the limit $|\omega_0| \to 0$. Since each component of the tachyonic spinors must satisfy the TKG equation, we infer that all the possible tachyonic wave equations that can be formulated in the framework of quantum mechanics describe the dynamics of subluminal particles with spacelike momentum. This result has been validated for the Chodos equation. In this context it is proved that the wave packet solution of this equation is characterized by a subluminal group velocity that tends asymptotically to zero as the corresponding classical particle velocity tends to infinitely large values. In other words, the tachyonic wave packet tends to behave like a localized wave for enough small energy values. Such behaviour could be explaining the long-standing controversy concerning the time spent by a massive particle to cross a classically forbidden barrier, i.e. the tunnelling time [30]. Theoretical investigations show that for enough large barriers the scattering velocity may become superluminal [9]. However, recently Ramos et al. have published an article [31] in which they state that although a peak appears at the output before the input even arrives, that does not mean anything travelled faster than light. This is due to the fact that *there is no law* connecting an incoming and an outgoing peak. We believe that this *law* may have something to do with the fact that within the potential barrier the particle behaves like a nearly localized PT wave packet. Inside the barrier, therefore, the particle behaves like a bradyon which satisfies the spacelike energy-momentum relation. This suggests that the scattering theory of a generic massive particle impinging a classically forbidden potential barrier should be completely revised.

**Declaration of Competing Interest**

The author declares that he has no known competing financial interests or personal relationships that could have appeared to influence the work reported in this paper.